\documentclass{article}

     \PassOptionsToPackage{numbers, compress}{natbib}
\usepackage{graphicx}
\usepackage[preprint]{neurips_2024}

\usepackage[utf8]{inputenc} % allow utf-8 input
\usepackage[T1]{fontenc}    % use 8-bit T1 fonts
\usepackage{hyperref}       % hyperlinks
\usepackage{url}            % simple URL typesetting
\usepackage{booktabs}       % professional-quality tables
\usepackage{amsfonts}       % blackboard math symbols
\usepackage{nicefrac}       % compact symbols for 1/2, etc.
\usepackage{microtype}      % microtypography
\usepackage{xcolor}         % colors

\title{Text2Tradition: From Epistemological Tensions to AI-Mediated Cross-Cultural Co-Creation}

\author{%
Pat Pataranutaporn$^1$ \quad Chayapatr Archiwaranguprok$^2$ \quad Phoomparin Mano$^3$ \quad \\
\textbf{Piyaporn Bhongse-tong}$^4$ \quad \textbf{Pattie Maes}$^1$ \quad \textbf{Pichet Klunchun}$^4$ \\
$^1$MIT Media Lab \quad $^2$University of the Thai Chamber of Commerce \quad \\$^3$Creatorsgarten \quad $^4$Pichet Klunchun Dance Company\\
}

\begin{document}

\maketitle

\begin{abstract}
% This paper introduces Text2Tradition, a system that attempts to bridge the epistemological gap between modern language processing and traditional dance knowledge. By leveraging large language models (LLMs) and rule-based traditional choreographic elements, we explore the tensions that arise when translating between these distinct knowledge systems. Our approach centers on the six traditional choreographic elements of No. 60 extracted from Mae Bot Yai, a Thai traditional dance repertoire. These elements represent an embodied, culturally-specific form of knowledge that has been passed down through generations. In contrast, LLMs embody a different type of knowledge - one that is data-driven, statistically derived, and often Western-centric. Text2Tradition attempts to create an interface between these two epistemologies. It uses LLMs to interpret textual inputs and manipulate pre-recorded 3D movements of a virtual dancer, resulting in novel choreographies. We argue that the tensions revealed by Text2Tradition highlight broader questions about the nature of knowledge, the role of embodiment in cultural practices, and the potential biases inherent in AI-mediated cultural interpretation. By critically examining these epistemological tensions, Text2Tradition contributes to ongoing discussions about the ethical and cultural implications of AI in preserving, interpreting, and potentially transforming traditional knowledge systems.

This paper introduces Text2Tradition, a system designed to bridge the epistemological gap between modern language processing and traditional dance knowledge by translating user-generated prompts into Thai classical dance sequences. Our approach focuses on six traditional choreographic elements from No. 60 in Mae Bot Yai, a revered Thai dance repertoire, which embodies culturally specific knowledge passed down through generations. In contrast, large language models (LLMs) represent a different form of knowledge—data-driven, statistically derived, and often Western-centric. This research explores the potential of AI-mediated systems to connect traditional and contemporary art forms, highlighting the epistemological tensions and opportunities in cross-cultural translation. Text2Tradition not only preserves traditional dance forms but also fosters new interpretations and cultural co-creations, suggesting that these tensions can be harnessed to stimulate cultural dialogue and innovation.
\end{abstract}

\section{Introduction}
In our diverse global landscape, the intersection of differing epistemologies often creates cultural tensions\cite{crosston2020cyber, williams2021enhancing, adams2021can}. Considering AI, LLMs promise universal translation of knowledge systems, yet exhibit biases and oversimplifications, particularly regarding underrepresented non-Western traditions \cite{maitra2020artificial}. This raises critical questions about epistemological representation in AI and cross-cultural knowledge integration. Rather than attempting to eliminate inevitable biases, we should focus on acknowledging, managing, and utilizing them. This paper presents a perspective on how epistemological tensions can be harnessed to create a symbiotic loop fostering cultural development and evolution, with each epistemology highlighting others' limitations and raising new questions about our approach to our worldviews.

Our work Text2Tradition presents a case study of an mediated interface between two artifacts: Thai traditional choreographic knowledge and LLM system. This study highlights the interplay between these distinct epistemologies. By examining user interactions with this tension between the two subjects, we explore how this interaction can lead to novel co-creation and deeper reflection on the complexities of cultural translation. Our aim is to foreground both the potentials and challenges of representing and engaging with diverse forms of embodied knowledge in computational systems and societies.

\begin{figure}
    \centering
    \includegraphics[width=1\linewidth]{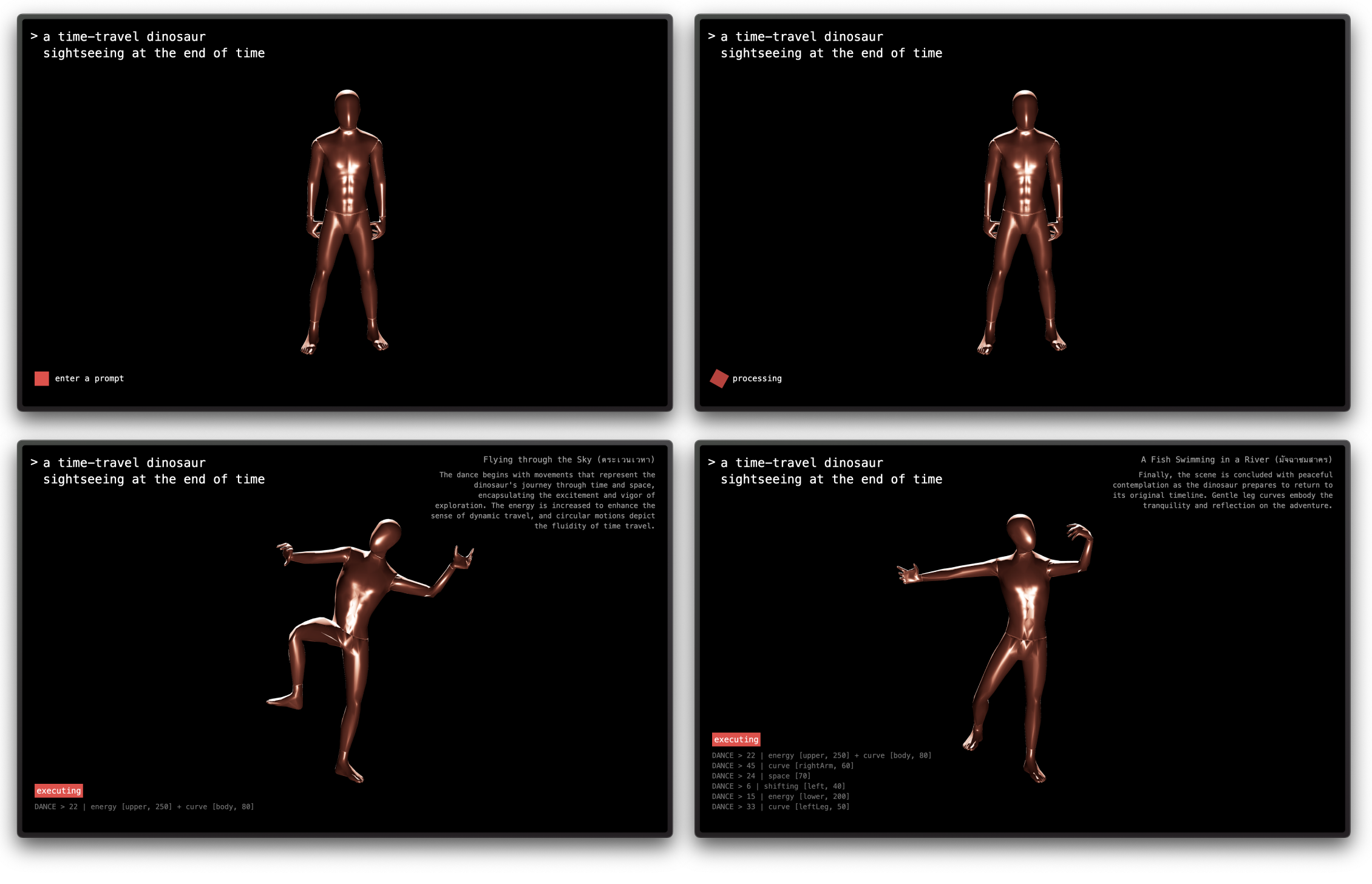}
    \caption{Web interface of text2tradition, encouraging new cultural co-creation through the reflection and interpretations between epistemological tensions of Thai dance and LLM dataset}
    \label{fig:screenshot}
\end{figure}

\section{Background and Related Work}

\subsection{Dance and Computation}
As a living embodiment of cultural heritage\cite{iacono2016beyond, aristidou2022safeguarding} encoding convolutional histories and contexts of its society, the choreographic intelligence encoded in traditional dance form serves as a compelling medium through which to explore these epistemological tensions. Prior to the consideration of cross-cultural translation, the form of dances themselves already raised important questions about the nature of creativity, embodiment, and cultural preservation in the digital age. Computational choreography has evolved significantly since Merce Cunningham's pioneering work with 3D computer graphics in the 1990s \cite{schiphorst1993case, schiphorst2013merce}. The field now encompasses a wide range of approaches, including motion capture \cite{carlson2015moment, oulasvirta2013information}, virtual environments \cite{giannachi2004virtual, farley2002digital}, wearable computing \cite{ladenheim2020live}, and AI-generated dance \cite{liu2024exploring, zhuang2022music2dance, Chan_2019_ICCV}. These technologies have not only expanded the creative possibilities for dance composition but also opened new avenues for documenting and preserving traditional dance forms as cultural heritage \cite{aristidou2022safeguarding, iacono2016beyond}. These approaches raise intriguing questions about the nature of creativity, the relationship between human and machine-generated choreography, and the potential for AI to bridge cultural and temporal divides in dance. However, the specific challenge of using text-based AI models to interpret and generate culturally specific dance movements remains largely unexplored.

\subsection{Cultural Context}
Thai traditional dance, or Natasin, has roots in the royal courts of ancient Siam, with a history spanning centuries~\cite{rutnin}, characterized by graceful, fluid movements, elaborate costumes, and complex hand gestures, each element carries specific meanings and symbolism~\cite{virulak1986status}. Central to Thai traditional dance is "Mae Bot Yai" (The Greater Fundamentals), a set of fundamental poses and movements \cite{skarreimagining, yamakup, brandon} that serve as a foundational alphabet for dance composition. A key characteristic of this art form is that each pose encodes flexible meanings, with a dance repertoire essentially being a composition and morphing of these alphabets~\footnote {For instance, the pose "Pisamai Riang Mon" conveys meanings related to love and relationships. Choreographers carefully select and subtly modify these poses to express specific narratives within their compositions, creating a nuanced language of movement.}

However, critics have noted a gradual loss of inner interpretability and dynamism in Thai traditional dance, arguing that movements have become self-stereotyped~\cite{prelim}. As a result, some contend that the dance has not significantly evolved since the early 20th century~\cite{pedagogy}, raising questions about cultural preservation versus artistic innovation. From this, a countermovement has recently emerged \cite{ngean2014choreographic}, aiming to reinvigorate traditional Thai dance by fostering innovation within its framework. Thai choreographer Pichet Klunchun, has been a significant contributor to this movement. Klunchun's project "No. 60," which serves as a foundation for our research, represents a culmination of his two-decade-long exploration of Thai classical dance \cite{skarreimagining}, alluding to the possibility of expanding beyond the traditional 59 movements of Mae Bot Yai. Klunchun distilled the Mae Bot Yai into six core principles dubbed as "six elements", including: Energy, Circles \& Curves, Axis Points, Synchronous Limbs, External Body Spaces, Shifting Relations.

\section{Methodology}
\subsection{Cyber Subin}
Cyber Subin\footnote{https://www.media.mit.edu/projects/cyber-subin/overview/}, a dance performance featuring co-dance between human dancers and generative virtual characters, builds upon the No. 60 project. It encodes the six elements identified in No. 60 into an algorithmic rule-based system \cite{moco}, which morphs prerecorded 3D virtual characters. The computational component, \textit{OpenCyberDance}\footnote{https://github.com/mitmedialab/opencyberdance}, responsible for virtual character generation and translation, is implemented as a web application using Vue and Three.js for 3D rendering. By digitally encoding traditional dance knowledge into (1) interactive 3D characters and (2) an algorithmic system, Cyber Subin opens new avenues for the evolution of Thai traditional dance in the digital landscape.

\subsection{Text2Tradition}
\begin{figure}
    \centering
    \includegraphics[width=1\linewidth]{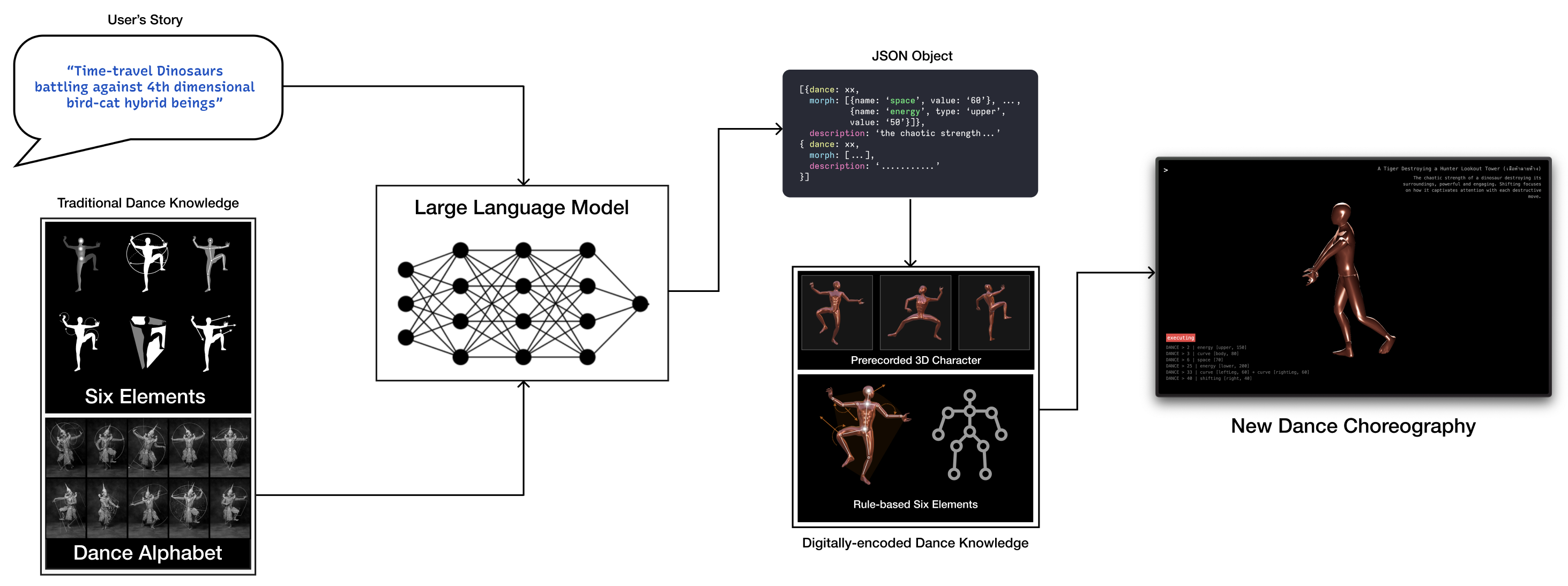}
    \caption{Dance Generation Process in Text2Tradition}
    \label{fig:diagram}
\end{figure}

Text2Tradition expands on Cyber Subin by shifting focus from reimagining "dance movement" to "dance repertoire." Recognizing Thai dance as a composition of movement alphabets, we explore contemporary interpretations of this art form. The platform enables users to input a story prompt (e.g., "A swan dancing"), which the system translates into a repertoire of Mae Bot Yai movements (such as "A Swan Gliding," "Flying Through the Sky," "A Dragon Playing in the Water"). Each movement can be further refined using the six elements system, for instance, lowering lower body energy to emphasize a swan's gentle elegance. Text2Tradition is a publicly accessible web platform designed to foster broader engagement in cultural co-creation. The dance generation process is illustrated in Figure~\ref{fig:diagram}.

\subsection{Technical Implementation}
The implementation of the Text2Tradition platform comprises three main components:
\begin{enumerate}
    \item \textbf{3D Character Library}: Unlike Cyber Subin, which used 3D recordings of existing dance repertoires (e.g., Kukpat and Yok Rob), Text2Tradition utilizes Mae Bot Yai movements as a dance alphabet. We conducted 3D motion-capture sessions with Thai traditional dance experts to record all 59 Mae Bot Yai movements. This motion data was applied to a 3D character, creating .glb files, which are then optimized using glTF transform for efficient web serving.

    \item \textbf{Dance Generator}: The core of Text2Tradition is a dance generation system utilizing LLM (GPT-4o) to interpret user prompts. It considers details of each Mae Bot Yai movement and incorporates the Six Elements principle as contextual data. The generator produces a JSON object based on the provided schema comprising an ordered array of Mae Bot Yai movements, with each potentially incorporating six elements refinements for nuanced adjustments aligning with the input story.

    \item \textbf{Web Interface}: Built on the \textit{OpenCyberDance} technological stack, the user-facing component is a web application using Vue and Three.js. It allows users to input story prompts, view generated dance sequences, and inspect the generator's interpretation. When a user runs a prompt, the dance generator is initiated. The resulting JSON object is then rendered on the web interface, where the renderer engine and rule-based six element system are implemented.
\end{enumerate}

\section{Results \& Discussion}
Text2Tradition highlights the potential of collaborative, interdisciplinary approaches in AI-mediated cultural exploration. This section presents three sample prompts and results from Text2Tradition, representing different types of stories and interpretations: (1) Star Wars: A New Hope, retold as a Thai Traditional Dance, (2) Lalisa dancing for a TikTok video, and (3) A steampunk adaptation of Stravinsky's The Rite of Spring. By embracing the interplay between AI systems and cultural contexts, it transforms perceived limitations into opportunities for dialogue and innovation. Considering dance generation, the process unfolds in three interconnected layers: the creative "translation" between cultures, the dynamic interaction of diverse epistemologies, and the resulting co-creation of new cultural expressions. 

\begin{figure}
    \centering
    \includegraphics[width=1\linewidth]{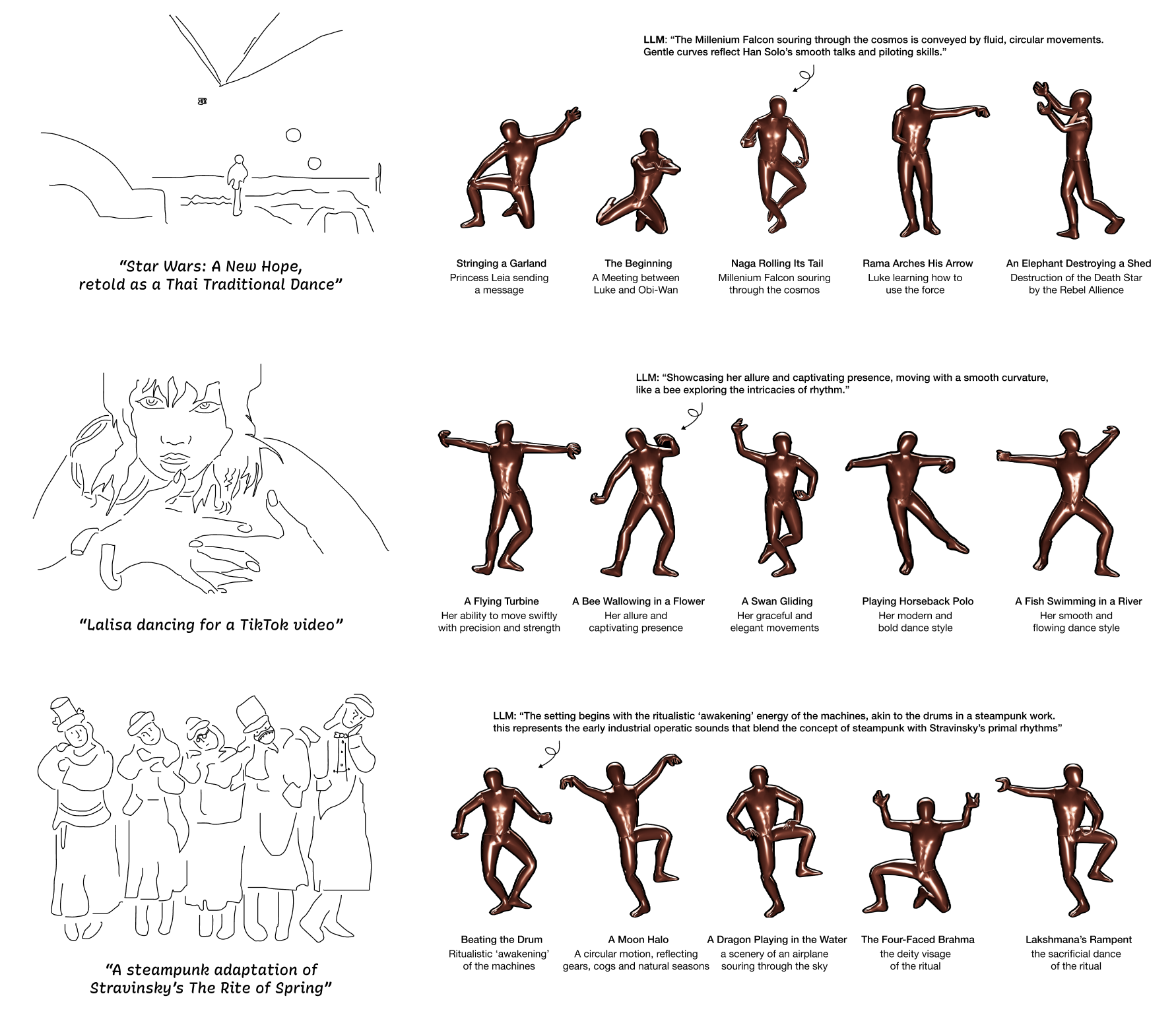}
    \caption{Three Generated Dances from Three Contemporary Tales}
    \label{fig:stories}
\end{figure}

\subsection{Narratives}
As depicted in Figure~\ref{fig:stories}, (1) is arguably the most straightforward, where the input is a linear story translated into another form of representation. The generated data is temporally respective to the storyline, with each pose representing significant moments of the story, both imitating characters (such as Princess Leia and Luke) and events (Destruction of the Death Star). The prompt of (2) is more conceptual, lacking a definite temporal storyline. In this and similar cases, the system usually generates a dance where each pose represents a metaphor of different aspects of the concept; in (2), each pose deconstructs Lalisa's characteristics and dance style. (3) is a blend between a linear story and a conceptual idea/theme, resulting in a new conceptual input. The result reflects this characteristic, with the generated output mixing parts of the storyline (such as "the sacrificial dance of the ritual") and conceptual representation ("ritualistic 'awakening' of the machine").

\subsection{Metaphors and Representations}

% The dance representations generated by Text2Tradition reflect multiple layers of conceptual metaphors~\cite{Black1981}. The primary layer involves the direct representation of narrative elements, translating persons, places, events, and abstract ideas into dance forms. A more nuanced secondary layer emerges from the intersection of cultural encodings and conceptual representations. We observe that as different cultures encode concepts uniquely in their artistic "alphabets," the collision of ideas in this AI-mediated space creates novel mappings of abstract metaphors~\cite{10.1145/3544548.3580700}. This process yields fresh understandings and meanings. For instance, in the Star Wars example, the destruction of the Death Star is depicted as an elephant destroying a shed. This multifaceted metaphor operates on several levels: the elephant pose embodies the massive scale of the Death Star, represents the collective strength of the Rebel forces, and mirrors the spectacular nature of the explosion within traditional contexts. This unexpected yet rich metaphor imbues the event with cultural significance while conveying the idea of a larger, more powerful force overcoming a seemingly impregnable structure. These examples demonstrate how Text2Tradition's output can serve as a bridge between disparate cultural and conceptual frameworks, offering new perspectives on familiar narratives through layered, culturally-informed physical metaphors.

Text2Tradition generates dance representations that reflect multiple layers of conceptual metaphors~\cite{Black1981}. The primary layer directly translates narrative elements into dance forms, while a secondary layer emerges from the intersection of cultural encodings and conceptual representations. As different cultures encode concepts uniquely in their artistic "alphabets," this AI-mediated space creates novel mappings of abstract metaphors~\cite{10.1145/3544548.3580700}, yielding fresh understandings. In the Star Wars example, the Death Star's destruction is depicted as an elephant destroying a shed. This metaphor may embodies the Death Star's scale, represents the Rebel forces' strength, and mirrors the explosion's spectacle within traditional contexts. This rich metaphor imbues the event with cultural significance while conveying the idea of a powerful force overcoming an impregnable structure. Text2Tradition thus bridges disparate cultural and conceptual frameworks, offering new perspectives on familiar narratives through layered, culturally-informed physical metaphors.

\subsection{Lost in Translation?}
While information loss is inherent in any translation process, including our system with its 59 Mae Bot Yai movements and six elements, viewing this solely as "lost in translation" represents a narrow, unidirectional perspective. In the ecosystem of this kind, the original input is not lost but complemented by the translation process. The resulting dance sequence should be seen not as an imperfect translation, but as a new creation inspired by and in dialogue with the original story. This shifts our focus from "what might be lost?" to "what is gained?" through creative interpretation.

Reframing the discussion, the cultural context and substitutions in the translated product highlight the system's constraints. However, these limitations may foster novel interpretations and unexpected connections, potentially revealing new dimensions that reflect the interplay of three forming epistemologies of the subjects in the translation process: the input story, the LLM translator, and the traditional dance form.

\subsection{Interpretation and Epistemological Interplay}
The dance generation process involves three key aspects of interpretation: (1) creating new stories that transcend traditional contexts, (2) selecting Mae Bot Yai movements based on their encoded meanings within the repertoire, and (3) morphing individual movements using the rule-based six elements to emphasize their significance relative to the narrative. This means that, the translation process is not a direct, one-fold translation, but a multistage process with each encoding its contextual factors. The interaction between these processes leads to a crossing of cultural contextual information, where each step loses and gains information, resulting in a dynamic system that yields new meanings and creative outputs.

Users also play a crucial role in this interplay, not merely as story creators and observers, but as active interpreters themselves. They should be encouraged to examine and rethink the implications of the translated product, understanding it not as an absolute answer, but as a signal for further examination, interpretation, and reconsideration.

\subsection{Toward Cross-Cultural Co-Creation}
The stagnation of cultural evolution has become an increasing concern in various domains~\cite{alivizatou2016intangible, peers2011making, aristidou2022safeguarding}, including Thai traditional dance \cite{pornrat}, where the art form risks losing relevance in contemporary society. Text2Tradition, like the work of Klunchun and other experimental preservationists~\cite{otero2016experimental}, aims to encourage engagement that fosters new perspectives and deepens cultural dialogue, allowing different epistemologies to learn from and evolve with each other. The technological approach employed in Cyber Subin and Text2Tradition has a crucial effect: the democratization of cultural interpretability. Rather than providing a static product, these systems offer dynamic tools for interpretation that can catalyze new co-creations between users and AI system. This approach enables broader societal participation in a multi-layered interpretative process, contributing to the ongoing evolution of both traditional dance forms and contemporary narrative expression. The result is a rich tapestry of cultural exchange and creativity, bridging the gap between preserved traditions and modern relevance.

\subsection{Future Directions}

\subsubsection{Expanding the Cultural Scope \& Enhancing Interactivity and Embodiment}

While our project focused specifically on Thai classical dance, the methodologies developed here could potentially be applied to other traditional art forms facing similar challenges of preservation and contemporary relevance. Future research could explore the application of this approach to diverse cultural traditions, contributing to a more globally inclusive development of AI in the arts. In addition, future iterations of the Text2Tradition system could explore more interactive and embodied forms of engagement. This might include the integration of motion capture technology to allow real-time interaction between human dancers and the AI system, or the development of augmented reality applications that could overlay AI-generated movements onto physical spaces.

\subsubsection{Considerations and Ethical Frameworks for AI in Cultural Domains}

AI systems like Text2Tradition enable cross-cultural co-creation but demand ongoing reevaluation to mitigate cultural oversimplification from Western-centric training data. While democratizing cultural interpretation is valuable, caution is necessary to prevent imposing dominant narratives or risking algorithmic colonization \cite{adams2021can}. Ethical frameworks for AI in cultural domains should emphasize inclusive design, diverse perspectives, and ownership rights, ensuring source communities maintain control over their cultural expressions. Future work should prioritize participatory approaches, involving communities in the co-creation process to enhance cultural authenticity and relevance.

% Participatory approaches, involving these communities in AI development and application, can mitigate risks of misrepresentation. Future work should focus on principles for responsible AI that balance innovation with cultural preservation and respect.

% In conclusion, Text2Tradition demonstrates the potential of AI to act as a bridge between traditional and modern forms of knowledge, translating the rich cultural heritage of Thai classical dance into new, dynamic expressions. By integrating the choreographic elements of Mae Bot Yai with the capabilities of large language models, the system not only preserves traditional art forms but also opens up avenues for cultural dialogue and innovation. This research underscores the importance of embracing epistemological tensions as opportunities for creative collaboration, ultimately contributing to the evolving landscape of cross-cultural artistic practices. As we continue to develop AI systems that engage with cultural heritage, it is crucial that we remain vigilant about the potential for algorithmic colonization and strive to create approaches that respect and empower diverse epistemologies. The future of AI in cultural domains lies not in the replacement of human expertise, but in the thoughtful integration of computational power with deep cultural knowledge and sensitivity.

In conclusion, Text2Tradition demonstrates AI's potential to bridge traditional and modern knowledge forms, translating Thai classical dance into new expressions. By integrating Mae Bot Yai with large language models, it preserves traditional art while fostering cultural dialogue and innovation. This research highlights the creative potential in epistemological tensions, contributing to cross-cultural artistic practices. As AI engages with cultural heritage, we must guard against algorithmic colonization and strive for approaches that respect diverse epistemologies. The future of AI in cultural domains lies in thoughtfully integrating computational power with cultural knowledge and sensitivity.

\bibliographystyle{IEEEtran}
\bibliography{reference}

\end{document}